\documentclass[10pt]{article}
\usepackage[dvips]{graphicx}
\usepackage{accents}
\begin{document}
\begin{center}
\textbf{{\large Statefinder Diagnostic for Dilaton Dark Energy}}
 \vskip 0.35 in
\begin{minipage}{4.5 in}
\begin{center}
{\small Z. G. Huang$^\dag$, X. M. Song \vskip 0.06 in \textit{
Department~of~Mathematics~and~Physics,
\\~Huaihai~Institute~of~Technology,~222005,~Lianyungang,~China
\\
$^\dag$zghuang@hhit.edu.cn}} \vskip 0.25 in {\small H. Q. Lu$^\ddag$
and W. Fang \vskip 0.06 in \textit{
Department~of~Physics,~Shanghai~University,~Shanghai,~China
\\
$^\ddag$alberthq$\_$lu@staff.shu.edu.cn}}
\end{center}
\vskip 0.2 in

{\small Statefinder diagnostic is a useful method which can differ
one dark energy model from the others. The Statefinder pair $\{r,
s\}$ is algebraically related to the equation of state of dark
energy and its first time derivative. We apply in this paper this
method to the dilaton dark energy model based on Weyl-Scaled induced
gravitational theory. We investigate the effect of the coupling
between matter and dilaton when the potential of dilaton field is
taken as the Mexican hat form. We find that the evolving trajectory
of our model in the $r-s$ diagram is quite different from those of
other dark energy models. \vskip 0.2 in \textit{Keywords:} Dark
energy; Statefinder; Dilaton; Mexican hat potential; Attractor.
}
\end{minipage}
\end{center}
\vskip 0.2 in
\par
Since the first observational data from SNe Ia[1] is issued in 1998,
exploring the nature of dark energy has been one of the most
challengeable problems in theoretical physics and modern
astrophysics. All data from SNe Ia[2] together with data from
WMAP5[3] and SDSS[4] strongly show us that, the Universe is
spatially flat with about one third of the critical energy density
being in non-relativistic matter and about two thirds of the
critical energy density being in a smooth component with large
negative pressure(dark energy), and is undergoing an accelerated
expansion phase. Of course, with the recent data on the galaxy power
spectrum from 2dF Galaxy Survey combined with CMB data[5], the
existence of dark energy(DE) can be proved without using the
supernova data at all[6].
\par So far, many models have been proposed to fit the observations including cosmological
constant $\Lambda$[7-11], quintessence[12-43], phantom[44-53],
holographic dark energy[54-60], Quintom[61-63], tachyon[64-73] and
Chaplygin gas[74,75] so on. The essential characteristics of these
dark energy models are contained in the parameter of its equation of
state, $p=\omega\rho$, where $p$ and $\rho$ denote the pressure and
energy density of dark energy, respectively, and $\omega$ is a state
parameter. Among these models, cosmological constant $\Lambda$ model
may be the simplest candidate. This constant term in Einstein field
equation can be regarded as an fluid with the equation of state
parameter $\omega=-1$. However, there are two serious problems with
the cosmological constant, namely the fine-tuning and the cosmic
coincidence. Firstly, in the framework of quantum field theory, the
vacuum expectation value is 123 order of magnitude larger than the
observed value of $10^{-47} GeV^4$. The absence of a fundamental
mechanism which sets the cosmological constant zero or very small
value is the cosmological constant "fine-tuning" problem. Secondly,
to explain in this way a constant vacuum energy density of $10^{-47}
GeV^4$, which is not only small but is also just the right value
that it is just beginning to dominate the energy density of the
Universe now, would require an unbelievable coincidence.
\par Quintessence model has been widely studied, and its state parameter $\omega$ which is
time-dependent, is greater than $-1$. In this paper, we regard
dilaton in Weyl-scaled induced gravitational theory as a
quintessence coupled with matter. It is well known that
scalar-tensor theories are the most natural extensions of general
relativity, in particular they contain local Lorentz invariance,
constancy of nongravitational constants and respect the weak
equivalence principle[76-84]. In our previous papers[85], we have
constructed a dilatonic dark energy model which belongs to
nonminimal quintessence[86-92], based on Weyl-scaled induced
gravitational theory. We found that when the dilaton field was not
gravitational clustered at small scales, the effect of dilaton can
not change the evolutionary law of baryon density perturbation, and
the density perturbation can grow from $z\sim10^3$ to $z\sim5$,
which guarantees the structure formation. We have also investigated
the property of the attractor solutions and concluded that the
coupling between dilaton and matter affects the evolutive process of
the Universe, but not the fate of the Universe.
\par With the remarkable increase in the accuracy of cosmological observational data during
the last few years and the appearance of more general models of dark
energy than a cosmological constant, advancing beyond quantities
Hubble parameter $H(t)\equiv\frac{\dot{a}}{a}$ and deceleration
parameter $q_0$ becomes a necessity. For this reason, Sahni et
al[93,94] propose a new geometrical diagnostic pair $\{r, s\}$ for
dark energy, which is called statefinder and can be expressed as
follows.
\begin{equation}r\equiv \frac{\dddot{a}}{aH^3},~~~~~~s\equiv\frac{r-1}{3(q-\frac{1}{2})}\end{equation}
where $r$ is a natural next step beyond $H$ and $q$. We can easily
see that this diagnostic is constructed from the $a(t)$ and its
derivatives up to the third order. So, the statefinder probes the
expansion dynamics of the universe through higher derivatives of the
expansion factor. By far, many models[95-100] have been
differentiated by this geometrical diagnostic method. Its important
property is that $\{r, s\} = \{1, 0\}$ is a fixed point for the flat
$\Lambda$CDM FRW cosmological model. Departure of a given DE model
from this fixed point is a good way of establishing the "distance"
of this model from flat $\Lambda$CDM. In this paper, we will
investigate the evolutive trajectory of our model in the $r-s$
diagram when the potential of dilaton field is taken as the Mexican
hat potential, and show the difference between our model and the
others, special $\Lambda$CDM.
\par The action of the Weyl-scaled induced gravitational theory is as follows:
\begin{equation}S=\int{d^4X\sqrt{-g}[\frac{1}{2}R(g_{\mu\nu})-\frac{1}{2}g^{\mu\nu}\partial_\mu\sigma\partial_\nu\sigma-V(\sigma)+L_{fluid}(\psi)}]\end{equation}
where
$L_{fluid}(\psi)=\frac{1}{2}g^{\mu\nu}e^{-\alpha\sigma}\partial_\mu\psi\partial_\nu\psi-e^{-2\alpha\sigma}V(\psi)$,
$\alpha=\sqrt{\frac{3M^2_p}{2\varpi+3}}$ with $\varpi>3500$[101]
being an important parameter in Weyl-scaled induced gravitational
theory, $\sigma$ is dilaton field, $g_{\mu\nu}$ is the Pauli metric
which can really represent the massless spin-two graviton and should
be considered to be physical metric[102], and $V(\sigma)$ is the
potential of dilaton field. The conventional Einstein gravity limit
occurs as $\sigma\rightarrow 0$ for an arbitrary $\varpi$ or
$\varpi\rightarrow\infty$ with an arbitrary $\sigma$. When
$V(\sigma)=0$, it will result in the Einstein-Brans-Dicke theory.
\par By varying action(2) and working in FRW universe, we obtain the field
equations of Weyl-scaled induced gravitational theory:
\begin{equation}H^2=\frac{1}{3M^2_p}(\rho_m+\rho_\sigma)\end{equation}
\begin{equation}\dot{H}=-\frac{1}{2M^2_p}(\rho_m+\rho_\sigma+p_\sigma)\end{equation}
\begin{equation}\dot{\rho}_m+3H\rho_m=\frac{1}{2}\alpha\dot{\sigma}\rho_m\end{equation}
\begin{equation}\dot{\rho}_\sigma+3H\dot{\sigma}^2=\frac{1}{2}\alpha e^{-\alpha\sigma}\rho_m\end{equation}
where $\rho_m$ is dark matter energy density, $\rho_\sigma$ is
dilaton dark energy energy density and radiation is neglected. The
effective energy density and pressure of dilaton dark energy can be
expressed as follows
\begin{equation}\rho_\sigma=\frac{1}{2}\dot{\sigma}^2+V(\sigma)\end{equation}
\begin{equation}p_\sigma=\frac{1}{2}\dot{\sigma}^2-V(\sigma)\end{equation}
For matter $p_m=0$, we get $\rho_m\propto
\frac{e^{\frac{1}{2}\alpha\sigma}}{a^3}$ from Eq.(5). Using Eq.(3)
and the $e$-$folding$ transformation $N=lna$, we have
\begin{equation}H=H_i[\frac{\frac{1}{2}\dot{\sigma}^2+V(\sigma)}{\rho_{c,i}}+\Omega_{m,i}e^{-\frac{1}{2}\alpha\sigma}e^{-3N}]^{\frac{1}{2}}\end{equation}
where $H_i^2=\frac{\rho_{c,i}}{3M^2_p}$, $\rho_{c,i}$ is the
critical energy density of the universe at initial time $t_i$.
$H_i$, $\Omega_{m,i}$, denote the Hubble parameter, matter energy
density parameter at initial time $t_i$ respectively.
\par Using the definition of Eq.(1) and Eqs.(3-6) , one can find that
\begin{equation}r=1+3\frac{\dot{H}}{H^2}+\frac{\ddot{H}}{H^3}
\\=1-\frac{3}{2}\Omega_\sigma\omega'_\sigma+\frac{9}{2}\omega_\sigma\Omega_\sigma(1+\Omega_\sigma)-\frac{3}{4}\alpha\sigma'(1-\Omega_\sigma)
-\frac{3}{4}\frac{\alpha
e^{-\alpha\sigma}}{H}(1-\Omega_\sigma)(1+\omega_\sigma)\end{equation}
\begin{equation}q=-1-\frac{\dot{H}}{H^2}=\frac{1}{2}(1+3\omega_\sigma\Omega_\sigma)\end{equation}
\begin{equation}s\equiv\frac{r-1}{3(q-\frac{1}{2})}=1+\omega_\sigma-\frac{1}{3}\frac{\omega'_\sigma}{\omega_\sigma}-
\frac{\alpha\sigma'}{6\omega_\sigma}\frac{1-\Omega_\sigma}{\Omega_\sigma}-\frac{1}{6}\frac{\alpha
e^{-\alpha\sigma}}{H}\frac{1-\Omega_\sigma}{\Omega_\sigma}\frac{1+\omega_\sigma}{\omega_\sigma}\end{equation}
where a prime denotes the derivative with respect to the
$e$-$folding$ time $N=lna$ and
$\Omega_i\equiv\frac{\rho_i}{3M^2_pH^2}$ for $i=m$ and $\sigma$.
\par Now, let us consider the Mexican hat potential
$V(\sigma)=\frac{\mu}{4}(\sigma^2-\varepsilon^2)^2+V_0$ where $\mu$,
$\varepsilon$ and $V_0$ are all constant. For this type of Mexican
hat potential, it has two extremum points in the range
$\sigma\geq0$: a minimum at $\sigma=\varepsilon$ and a maximum at
$\sigma=0$. The non-conventional parameter $V_0$ in this potential
moves the potential up and down, which is equivalent to adding a
cosmological constant to the usual Mexican hat potential. We show
the features of the Mexican hat potential mathematically in Fig.1.
\vskip 0.3 in
\begin{center}
\begin{minipage}{0.6\textwidth}
\includegraphics[scale=1]{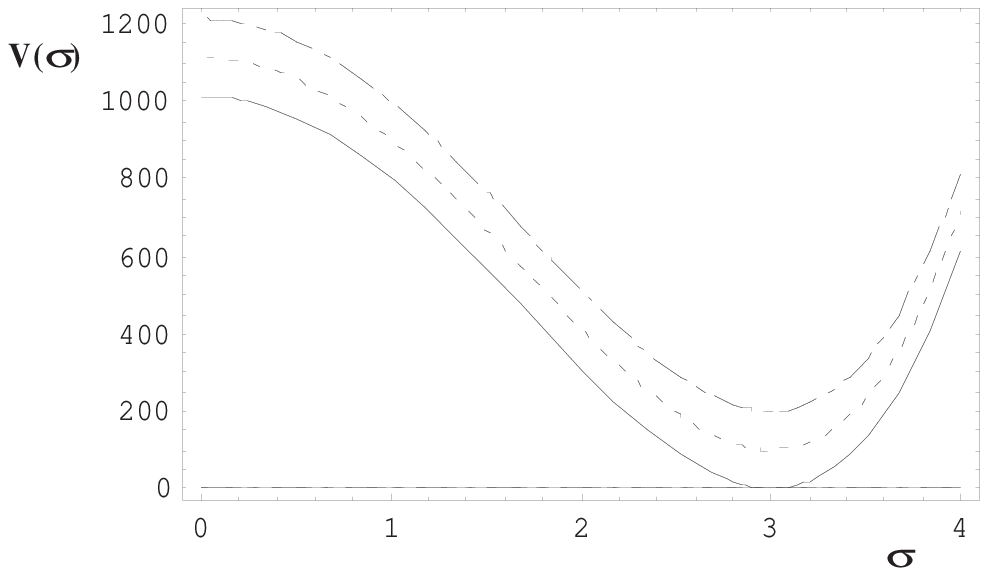}
{\small Fig.1 The Mexican hat potential
$V(\sigma)=\frac{\mu}{4}(\sigma^2-\varepsilon^2)^2+V_0$. We set
$V_0=0$(real line), 100(dot line), 200(dot-dashed line).}
\end{minipage}
\end{center}

\vskip 0.3 in
\begin{center}
\begin{minipage}{0.6\textwidth}
\includegraphics[scale=1]{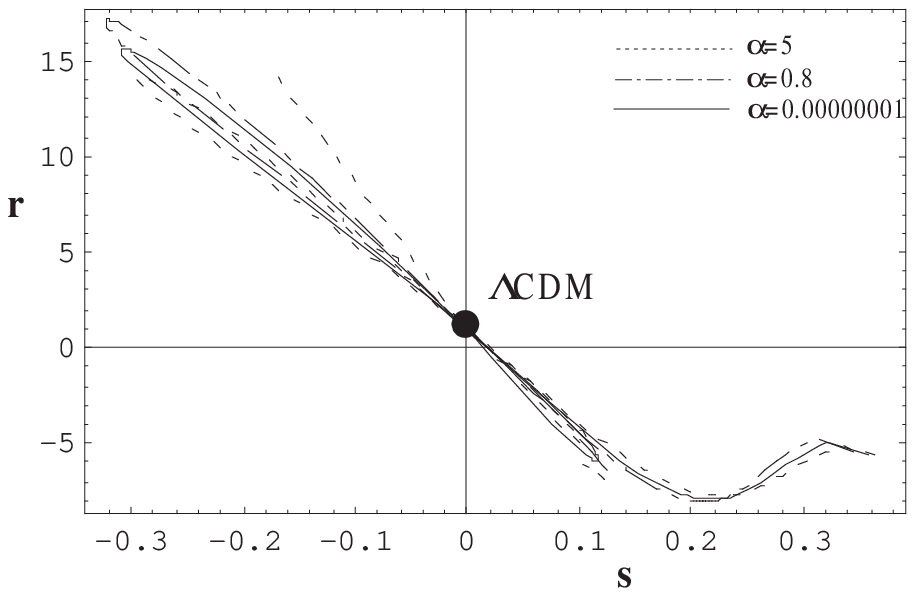}
{\small Fig.2 The $r-s$ diagram of Mexican hat potential
$V(\sigma)=\frac{\mu}{4}(\sigma^2-\varepsilon^2)^2+V_0$. Curves
$r(s)$ evolves in the $e$-fold time interval $N\in[0,~0.878]$. The
black dot corresponds the fixed point of $\Lambda$CDM,
$\{r=1,~s=0\}$. $\alpha$ denoting the intensity of coupling between
dilaton and matter, is set for $\alpha=0.00000001$(real line),
$\alpha=0.8$(dot-dashed line) and $\alpha=5$(dot line) respectively.
}
\end{minipage}
\end{center}

\vskip 0.15 in
\begin{center}
\begin{minipage}{0.6\textwidth}
\includegraphics[scale=1]{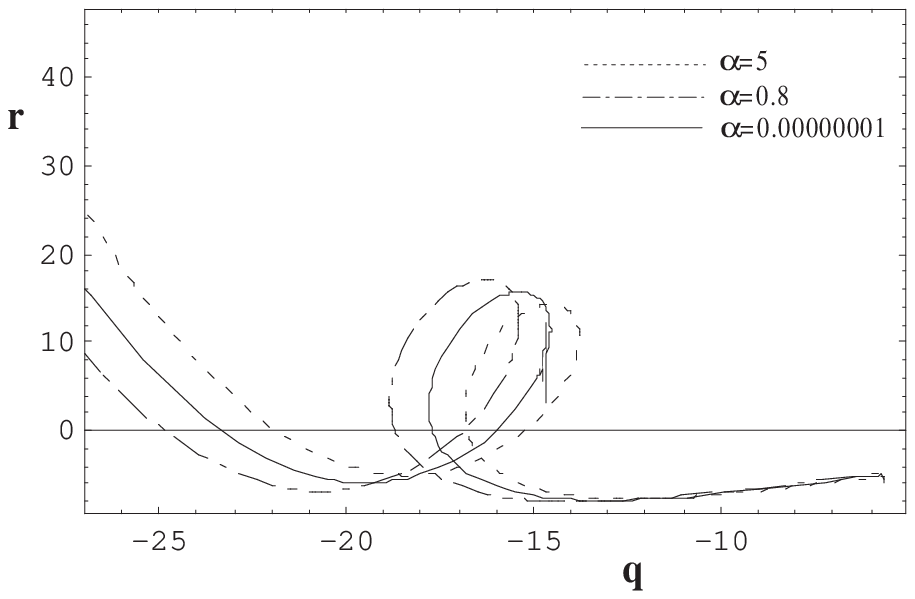}
{\small Fig.3 The diagram $r-q$ of Mexican hat potential
$V(\sigma)=\frac{\mu}{4}(\sigma^2-\varepsilon^2)^2+V_0$ when we set
 $\alpha=0.00000001$(real line), $\alpha=0.8$(dot-dashed line) and
$\alpha=5$(dot line) respectively.}
\end{minipage}
\end{center}

\vskip 0.3 in
\begin{center}
\begin{minipage}{0.6\textwidth}
\includegraphics[scale=1]{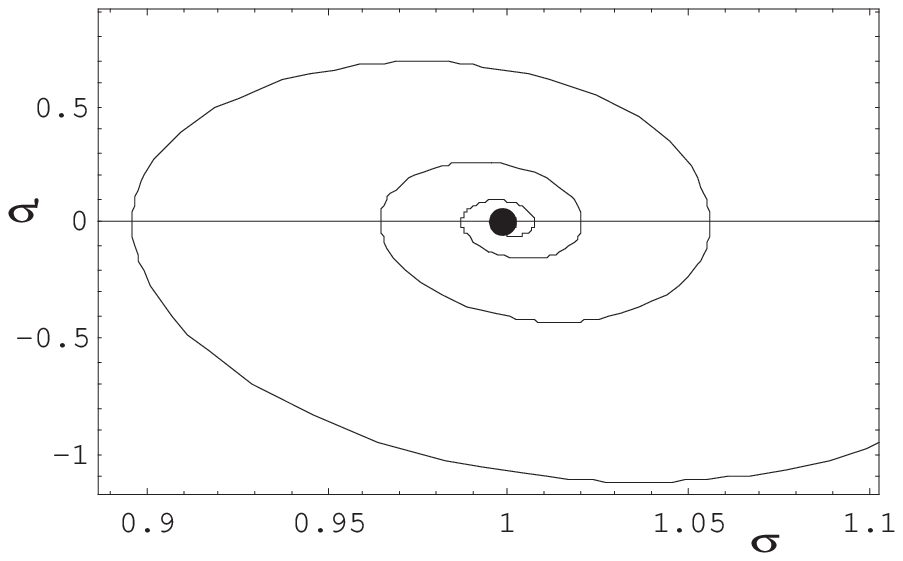}
{\small Fig.4 The phase portrait of attractor in the Mexican hat
potential for $\alpha=0.00000001$. The trajectory of attractor is a
stable spiral.}
\end{minipage}
\end{center}
\vskip 0.15 in
\par In Fig.2, the $\{r,s\}$ phase portrait is shown numerically.
When the coupling parameter $\alpha$ is set 5(dot line),
0.8(dot-dashed line) and 0.00000001(real line), the evolutive
trajectories of $r(s)$ are very similar. This means that the
intensity of coupling between dilaton and matter changes the
evolutive trajectory of $r(s)$ weakly. This result is consistent
with a conclusion obtained from our previous paper[14]: the coupling
between dilaton and matter affects the evolutive process of the
Universe, but not the fate of the Universe. We can easily see that
the trajectory of $r(s)$ will pass the fixed point $\{r=1,~s=0\}$ of
$\Lambda$CDM in the future and is different from the other dark
energy models. Fig.3 shows the evolutive behavior of parameter $r$
with respect to deceleration parameter in the range of $e$-$folding$
time $N\in[0,~0.92]$. The evolutive behavior for different coupling
parameter $\alpha$ will tend to the same one. Fig.4 shows that the
shape of the evolutive trajectory of $\sigma'-\sigma$ is a stable
spiral corresponding to the equation of state $\omega=-1$ and the
dilaton dark energy density parameter $\Omega=1$, which are
important features for a dark energy model that can meet the current
observations.
\par In summary, we apply the statefinder diagnostic
to the dilaton dark energy model based on the Weyl-scaled induced
gravitational theory with Mexican hat potential
$V(\sigma)=\frac{\mu}{4}(\sigma^2-\varepsilon^2)^2+V_0$. The effect
of coupling between dilaton and matter on the evolutive trajectory
of $r(s)$ with respect to the $e$-$folding$ time $N=lna$, is
investigated in this paper. First, we get the attractor solution is
a stable spiral. Second, according to the numerical results, we get
the coupling between dilaton and matter changes the evolutive
behavior of $r(s)$ very weakly and the trajectories of $r(s)$ for
different coupling parameter $\alpha$ will always pass the fixed
point $\{r=1,~s=0\}$ corresponding to $\Lambda$CDM model. At last,
we find that the evolutive trajectory of $r(s)$ forms a swirl before
reaches attractor and is quite differen from those of other dark
energy models[93-100].

\vskip 0.25 in
\begin{flushleft}\textbf{Acknowledgements}\end{flushleft}
This work is partially supported by National Nature Science
Foundation of China under Grant No.10573012, National Nature Science
Foundation of HHIT under Grant No.Z2007022 and National Nature
Science Foundation of Jiangsu Province under Grant No.07KJD140011.

\begin{flushleft}{\noindent\bf References}
 \small{

\item {1.}{ A. G. Riess, \textit{Astron. J}\textbf{116}, 1009(1998).}
\item {2.}{ A. G. Riess et al., \textit{Astrophys. J}\textbf{607}, 665(2004);
\\\hspace{0.15 in}S. Perlmutter et al., \textit{Astrophys. J}\textbf{517}, 565(1999);
\\\hspace{0.15 in}N. A. Bahcall et al., \textit{Science}\textbf{284}, 1481(1999).}
\item {3.}{ G. Hinshaw et al., arXiv: 0803.0732;
\\\hspace{0.15 in}M. R. Nolta, et al., arXiv: 0803.0593.}
\item {4.}{ M. Tegmark, et al., \textit{Phys. Rev. D}\textbf{69}, 103510(2004);}
\item {5.}{ D. N. Spergel et al., \textit{Astrophys. J. Suppl}\textbf{148}, 175(2003).}
\item {6.}{ W.J. Percival et al., \textit{Mon. Not. Roy. Astron. Soc}\textbf{337}, 1068(2002).}
\item {7.}{ S. Weinberg \textit{Rev. Mod. Phys.}\textbf{61}, 1(1989).}
\item {8.}{ V. Sahni and A. Starobinsky, \textit{Int. J. Mod. Phys. D}\textbf{9}, 373(2000).}
\item {9.}{ S. M. Carroll, \textit{Living Rev. Rel.}\textbf{4}, 1(2001).}
\item {10.}{P. J. E. Peebles and B. Ratra, \textit{Rev. Mod. Phys.}\textbf{75}, 559(2003).}
\item {11.}{T. Padmanabhan, \textit{Phys. Rept.}\textbf{380}, 235(2003).}
\item {12.}{L. Amendola, M. Quartin, S. Tsujikawa and I. Waga, \textit{Phys.Rev.D}\textbf{74}, 023525(2006.}
\item {13.}{E. Elizalde, S. Nojiri and S. D. Odintsov, arXiv:hep-th/0405034.}
\item {14.}{S. Nojiri, S. D. Odintsov and M. Sasaki, \textit{Phys.Rev. D}\textbf{70} 043539(2004).}
\item {15.}{S. Nojiri and S. D. Odintsov \textit{Phys. Lett. B}\textbf{639}, 144(2006).}
\item {16.}{B. Boisseau et al., \textit{Phys. Rev. Lett}\textbf{85}, 2236(2000), arXiv:gr-qc/0001066.}
\item {17.}{G. Esposito-Farese and D. Polarski, \textit{Phys. Rev. D}\textbf{63}, 063504(2001), arXiv:gr-qc/0009034.}
\item {18.}{X. Zhang, \textit{Phys. Lett. B}\textbf{611}, 1(2005), arXiv:astro-ph/0503075.}
\item {19.}{M. R. Setare, \textit{Phys. Lett. B}\textbf{642},1(2006), arXiv:hep-th/0609069.}
\item {20.}{V. Faraoni and M. N. Jensen, gr-qc/0602097.}
\item {21.}{C. Wetterich \textit{Nucl. Phys. B}\textbf{302}, 668(1988).}
\item {22.}{E. J. Copeland, M. Sami and S. Tsujikawa, arXiv:hep-th/0603057.}
\item {23.}{P. G. Ferreira and M. Joyce \textit{Phys. Rev. D}\textbf{58}, 023503(1998).}
\item {24.}{J. Frieman, C. T. Hill, A. Stebbinsand and I.Waga, \textit{Phys. Rev. Lett}\textbf{75}, 2077(1995).}
\item {25.}{P. Brax and J. Martin, \textit{Phys. Rev. D}\textbf{61}, 103502(2000).}
\item {26.}{T. Barreiro, E. J. Copeland and N. J. Nunes, \textit{Phys. Rev. D}\textbf{61}, 127301(2000).}
\item {27.}{I. Zlatev, L. Wang and P. J. Steinhardt \textit{Phys. Rev. Lett} \textbf{82}, 896(1999).}
\item {28.}{T. Padmanabhan, and T. R. Choudhury, \textit{Phys. Rev. D}\textbf{66}, 081301(2002).}
\item {29.}{A. Sen, \textit{JHEP} \textbf{0204}, 048(2002).}
\item {30.}{C. Armendariz-Picon, T. Damour and V. Mukhanov, \textit{Phys. Lett. B}\textbf{458}, 209(1999).}
\item {31.}{A. Feinstein, \textit{Phys. Rev. D}\textbf{66}, 063511(2002).}
\item {32.}{M. Fairbairn and M. H. Tytgat, \textit{Phys. Lett. B}\textbf{546}, 1(2002).}
\item {33.}{A. Frolov, L. Kofman and A. Starobinsky, \textit{Phys.Lett.B} \textbf{545}, 8(2002).}
\item {34.}{L. Kofman and A. Linde, \textit{JHEP}\textbf{0207}, 004(2004).}
\item {35.}{C. Acatrinei and C. Sochichiu, \textit{Mod. Phys. Lett. A}\textbf{18}, 31(2003).}
\item {36.}{S. H. Alexander, \textit{Phys. Rev. D}\textbf{65}, 0203507(2002).}
\item {37.}{T. Padmanabhan, \textit{Phys. Rev. D}\textbf{66}, 021301(2002).}
\item {38.}{A. Mazumadar, S. Panda and A. Perez-Lorenzana, \textit{Nucl. Phys. B}\textbf{614}, 101(2001).}
\item {39.}{S. Sarangi and S. H. Tye, \textit{Phys. Lett. B}\textbf{536}, 185(2002).}
\item {40.}{Z. G. Huang and H. Q. Lu, \textit{Int. J. Mod. Phys. D}\textbf{15}, 1501(2006).}
\item {41.}{Z. G. Huang, H. Q. Lu and W. Fang, \textit{Int. J. Mod. Phys. D}\textbf{16}, 1109(2007), arXiv:hep-th/0610018.}
\item {42.}{Z. G. Huang, X. H. Li and Q. Q. Sun, \textit{Astrophys. Space Sci.}\textbf{310},53(2007), arXiv:hep-th/0610019.}
\item {43.}{W. Fang, H. Q. Lu and Z.G. Huang, \textit{Class. Quant. Grav}\textbf{24}, 3799(2007).}
\item {44.}{ L. Amendola, S. Tsujikawa, and M. Sami, \textit{Phys. Lett. B}\textbf{632}, 155(2006).}
\item {45.}{L. Amendola, \textit{Phys. Rev. Lett.}\textbf{93}, 181102(2004).}
\item {46.}{R. Gannouji, D. Polarski, A. Ranquet and A. A. Starobinsky, arXiv:astro-ph/060628.}
\item {47.}{H. Q. Lu, \textit{Int. J. Mod. Phys. D}\textbf{14}, 355(2005), arXiv:hep-th/0312082.}
\item {48.}{H. Q. Lu, Z. G. Huang, W. Fang and P. Y. Ji, arXiv:hep-th/0504038.}
\item {49.}{W. Fang et al., \textit{Int. J. Mod. Phys. D}\textbf{15}, 199(2006), arXiv:hep-th/0409080.}
\item {50.}{X. Z. Li and J. G. Hao, \textit{Phys. Rev. D}\textbf{69}, 107303(2004).}
\item {51.}{T. Chiba, T. Okabe and M. Yamaguchi, \textit{Phys. Rev. D}\textbf{62}, 023511(2000).}
\item {52.}{P. Singh, M. Sami and N. Dadhich, \textit{Phys.Rev. D}\textbf{68}, 023522(2003).}
\item {53.}{S. M. Carroll, M. Hoffman and M. Trodden, \textit{Phys. Rev. D}\textbf{68}, 023509(2003).}
\item {54.}{M. Li, \textit{Phys. Lett. B}\textbf{603} 1(2004), arXiv:hep-th/0403127.}
\item{55.}{Q. G. Huang and M. Li, \textit{JCAP}\textbf{0408}, 013(2004).}
\item{56.}{M. Ito, \textit{Europhys. Lett}.\textbf{71}, 712-715(2005).}
\item{57.}{K. Ke and M. Li, \textit{Phys.Lett.B}\textbf{606}, 173-176(2005).}
\item{58.}{Q. G. Huang and M. Li, \textit{JCAP}\textbf{0503}, 001(2005).}
\item{59.}{Y. G. Gong, B. Wang and Y. Z. Zhang, \textit{Phys. Rev. D}\textbf{72}, 043510(2005).}
\item{60.}{X. Zhang, \textit{Int. J. Mod. Phys. D}\textbf{14}, 1597-1606(2005).}
\item{61.}{W. Hao, R. G. Cai and D. F. Zeng, \textit{Class.Quant.Grav}\textbf{22}, 3189(2005).}
\item{62.}{Z. K. Guo, Y. S. Piao, X. M. Zhang, Y.Z. Zhang, \textit{Phys.Lett. B}\textbf{608}, 177(2005).}
\item{63.}{B. Feng, arXiv:astro-ph/0602156.}
\item{64.}{A. Sen, \textit{JHEP} \textbf{0207}, 065(2002).}
\item{65.}{M. R. Garousi, \textit{Nucl. Phys. B}\textbf{584}, 284(2000).}
\item{66.}{M. R. Garousi, \textit{JHEP} \textbf{0305}, 058(2003).}
\item{67.}{E. A. Bergshoeff, M. de Roo, T. C. de Wit, E. Eyras and S. Panda,\textit{ JHEP} \textbf{0005}, 009(2000).}
\item{68.}{J. Kluson, \textit{Phys. Rev. D}\textbf{62}, 126003(2000).}
\item{69.}{G. W. Gibbons, \textit{Phys. Lett. B}\textbf{537}, 1(2002).}
\item{70.}{M. Sami, P. Chingangbam and T. Qureshi, \textit{Phys. Rev. D}\textbf{66}, 043530(2002).}
\item{71.}{M. Sami, \textit{Mod. Phys. Lett. A}\textbf{18}, 691(2003).}
\item{72.}{Y. S. Piao, R. G. Cai, X. m. Zhang and Y. Z. Zhang, \textit{Phys. Rev. D}\textbf{66},121301(2002).}
\item{73.}{L. Kofman and A. Linde, \textit{JHEP} \textbf{0207}, 004(2002).}
\item{74.}{M. K. Mak and T. Harko, \textit{Phys.Rev.D.} \textbf{71}, 104022(2005).}
\item{75.}{H. Q. Lu et al., arXiv:hep-th/0504038.}
\item{76.}{A. Riazuelo and J. P. Uzan, \textit{Phys. Rev. D}\textbf{66}, 023525 (2002).}
\item{77.}{C. M. Will, Theory and Experiments in Gravitational Physics(Cambridge University Press, England, 1993).}
\item{78.}{N. Bartolo and M. Pietroni, \textit{Phys. Rev. D}\textbf{61}, 023518 (2000), arXiv:hep-ph/9908521.}
\item{79.}{T. Damour and K. Nordtvedt,  \textit{Phys. Rev. D}\textbf{48}, 3436 (1993).}
\item{80.}{X. Chen and M. Kamionkowsky, \textit{Phys. Rev. D}\textbf{60}, 104036 (1999), arXiv:astro-ph/9905368.}
\item{81.}{C. Baccigaluppi, S. Matarrese, and F. Perrotta, \textit{Phys. Rev. D}\textbf{62}, 123510 (2000), arXiv:astro-ph/0005543.}
\item{82.}{L. Amendola, \textit{Month. Not. R. Astron. Soc.}\textbf{312}, 521(2000), arXiv:astro-ph/9906073.}
\item{83.}{L. Amendola,  \textit{Phys. Rev. Lett.}\textbf{86}, 196(2001), arXiv:astro-ph/0006300.}
\item{84.}{X. Chen, R.J. Scherrer, and G. Steigman, \textit{Phys. Rev. D}\textbf{63},123504(2001), arXiv:astro-ph0011531.}
\item{85.}{Z. G. Huang, H. Q. Lu and W. Fang, \textit{Class. Quant. Grav}\textbf{23}, 6215(2006), arXiv:hep-th/0604160.}
\item{86.}{J. P. Uzan, \textit{Phys. Rev. D}\textbf{59}, 123510 (1999), arXiv:gr-qc/9903004.}
\item{87.}{L. Amendola, \textit{Phys. Rev. D}\textbf{62}, 043511 (2000), arXiv:astro-ph/9908023.}
\item{88.}{F. Perrotta, C. Baccigalupi, and S. Matarrese, \textit{Phys. Rev. D}\textbf{61},023507(2000), arXiv:astro-ph/9906066.}
\item{89.}{G. Esposito-Far`ese and D. Polarski, \textit{Phys. Rev. D}\textbf{63}, 063504 (2001), arXiv:gr-qc/0009034.}
\item{90.}{L. Amendola, \textit{Phys. Rev. D}\textbf{60}, 043501 (1999), arXiv:astro-ph/9904120.}
\item{91.}{R. de Ritis, A.A. Marino, C. Rubano, and P. Scudellaro, \textit{Phys. Rev. D}\textbf{62}, 043506(2000), arXiv:hep-th/9907198.}
\item{92.}{O. Bertolami and P.J. Martins, \textit{Phys. Rev. D}\textbf{61}, 064007(2000), arXiv:gr-qc/9910056.}
\item{93.}{V. Sahni, T. D. Saini, A. A. Starobinsky and U. Alam, \textit{JETP Lett}\textbf{77}, 201(2003);}
\item{94.}{U. Alam, V. Sahni, T. D. Saini and A. A. Starobinsky, \textit{Mon. Not. Roy. ast. Soc}\textbf{344}, 1057(2003).}
\item{95.}{B. R. Chang, H. Y. Liu, L. X. Xu, C. W. Zhang and Y. L. Ping, \textit{JCAP}\textbf{0701}, 016(2007).}
\item{96.}{Gorini, A. Kamenshchik and U. Moschella, \textit{Phys. Rev. D}\textbf{67}, 063509(2003).}
\item{97.}{X. Zhang, \textit{Phys. Lett. B}\textbf{611}, 1(2005).}
\item{98.}{X. Zhang, \textit{Int. J. Mod. Phys. D}\textbf{14}, 1597(2005).}
\item{99.}{P. X. Wu and H. W. Yu, \textit{Int. J. Mod. Phys. D}\textbf{14}, 1873(2005).}
\item{100.}{X. Zhang, F. Q. Wu, J. F. Zhang, \textit{JCAP}\textbf{ 0601}, 003(2006).}
\item{101.}{C. M. Will, \textit{Living Rev. Rel.}\textbf{4}, 4(2001).}
\item{102.}{Y. M. Cho, \textit{Phys. Rev.Lett}\textbf{68}, 3133(1992).}
}
\end{flushleft}
\end{document}